\documentclass[%
 reprint,
showpacs,
nofootinbib,
 amsmath,amssymb,
 aps,
 prd
]{revtex4-1}

\usepackage{graphicx}% Include figure files
\usepackage{dcolumn}% Align table columns on decimal point
\usepackage{bm}% bold math
\usepackage{color}

\begin{document}

%\preprint{APS/123-QED}

\title{Energy-momentum tensors in linearized Einstein's theory and\\ massive gravity: The question of uniqueness}% Force line breaks with \\

\author{Ji\v{r}\'{i} Bi\v{c}\'{a}k}
\affiliation{%
 Institute of Theoretical Physics, Faculty of Mathematics and Physics, Charles University, V Hole\v{s}ovi\v{c}k\'{a}ch 2, 180 00 Praha 8, Czech Republic
}%
\affiliation{%
 Max Planck Institute for Gravitational Physics, Albert Einstein Institute, Am M\"{u}hlenberg 1, D-14476 Golm, Germany
}%

\author{Josef Schmidt}
\affiliation{
Department of Physics, Faculty of Nuclear Physics and Physical Engineering, Czech Technical University in Prague, B\v{r}ehov\'{a} 7, Praha 1, Czech Republic
}%


\begin{abstract}
The question of the uniqueness of energy-momentum tensors in the linearized general relativity and in the linear massive gravity is analyzed without using variational techniques. We start from a natural ansatz for the form of the tensor (for example, that it is a linear combination of the terms quadratic in the first derivatives), and require it to be conserved as a consequence of field equations. In the case of the linear gravity in a general gauge we find a four-parametric system of conserved second-rank tensors which contains a \textit{unique} symmetric tensor. This turns out to be the linearized Landau-Lifshitz pseudotensor employed often in full general relativity. We elucidate the relation of the four-parametric system to the expression proposed recently by Butcher \textit{et al.} ``on physical grounds'' in harmonic gauge, and we show that the results coincide in the case of high-frequency waves in vacuum after a suitable averaging. In the massive gravity we show how one can arrive at the expression which coincides with the ``generalized linear symmetric Landau-Lifshitz'' tensor. However, there exists another uniquely given simpler symmetric tensor which can be obtained by adding the divergence of a suitable superpotential to the canonical energy-momentum tensor following from the Fierz-Pauli action. In contrast to the symmetric tensor derived by the Belinfante procedure which involves the second derivatives of the field variables, this expression contains only the field and its first derivatives. It is simpler than the generalized Landau-Lifshitz tensor but both yield the same total quantities since they differ by the divergence of a superpotential. We also discuss the role of the gauge conditions in the proofs of the uniqueness. In the Appendix, the symbolic tensor manipulation software \textsc{Cadabra} is briefly described. It is very effective in obtaining various results which would otherwise require lengthy calculations.
\end{abstract}


\maketitle

\section{Introduction}

In relativistic field theories the most frequently used method of constructing conserved quantities is based on Noether's theorems. If one starts from the Lorentz invariant Lagrangian, one can employ its symmetries and form (in general asymmetric) a canonical energy-momentum tensor which can be symmetrized by adding the divergence of a superpotential. Alternatively, one may rewrite the Lagrangian in a manifestly covariant manner and obtain, by inducing the variation of the metric by infinitesimal coordinate transformations, the symmetric tensor directly.  

Although the quantities so derived are well established and physically sound, there may exist other second-rank tensors which may be useful just because they are \textit{conserved} as a consequence of the field equations. For example, they may turn out to control the time evolution of the Cauchy data more efficiently than ``conventional'' expressions (e.g. the fourth-rank Bel-Robinson tensor is used in the proofs of the global nonlinear stability of Minkowski space). In addition, physical theories exist, for which the field equations are not derivable from a variational principle; this is the case with, for example, the ``already unified theory'' of gravity and electromagnetism by Rainich \cite{rainich}.

In the following, we consider, within linearized Einstein's theory and massive gravity, various second-rank tensors involving fields and their first derivatives conserved as the consequence of the field equations involving their second derivatives. Although we fully acknowledge the significance of the expressions derived from the variational principle as mentioned above, we take the liberty to call sometimes such conserved second-rank tensors ``energy-momentum tensors'' or ``complexes'' being influenced, among others, by language used frequently in general relativity. 

The technique we use to construct expressions conserved modulo field equations goes back to the work of Fock.  At the 1962 Warsaw conference on general relativity \cite{fock} and in the second (revised) edition of his influential monograph, Fock \cite{fock2} summarized his work on the uniqueness of the energy-momentum tensors of the electromagnetic field, of incoherent matter, and of a perfect fluid in which a Lagrangian formalism is not used. The uniqueness was proven assuming the energy-momentum tensor is a symmetric tensor of the second order, formed with the field variables, and conserved as a consequence of field equations.

Inspired by Fock's work, one of us generalized this method to the wave fields described by the equations of the second order -- neutral scalar meson field, vector (Proca) field, linearized gravitational field and the gravitational field in full nonlinear Einstein's theory \cite{bicak}, \cite{bicak2}. As one proceeds to more complicated theories, some simplifying assumptions about the structure of the expressions for the energy-momentum are made. In the case of linearized Einstein's equations when the metric tensor, in a suitable coordinate system, can be written in the form $g_{ik} = \eta_{ik} + h_{ik}$, $|h_{ik}| \ll 1$, the basic assumption is that the energy-momentum tensor $T_{ik}$ depends on 20 independent quantities quadratic in $h_{ik,l}$; however, the symmetry of $T_{ik}$ was not assumed. Also, no specific gauge was chosen. It turned out that the resulting expression conserved modulo linearized Einstein's equations forms a four-parameter system. Among these, there is the linearized Einstein's complex; it is nonsymmetric and can be derived from a Lagrangian. However, there exists also a \textit{unique symmetric tensor} which does not follow from a variational principle. We show that it is the linearized Landau-Lifshitz pseudotensor frequently used in full general relativity and in approximation methods going beyond the linear theory (cf. e.g. \cite{MTW}, \cite{landau}, \cite{will}).

Recently, Butcher \textit{et al.} from the Cambridge Kavli Institute for Cosmology published a series of papers on ``localized energetics of linear gravity'' \cite{butcher}, \cite{butcher3}, \cite{butcher2}. By examining the transfer of energy and momentum between local matter and the gravitational field within the linearized theory, they constructed a symmetric energy-momentum tensor of linearized gravity which exhibits plausible physical properties and is quadratic in the first derivatives $h_{ik,l}$; however, the whole framework leads to the use of the harmonic gauge \cite{butcher}. Later the same authors extended their work to the study of the localized angular momentum of linearized gravity \cite{butcher3}. They also constructed a Fierz-Pauli Lagrangian for a massless spin-2 field and made it covariant by introducing the nonholonomic basis (tetrad) and connection which in general led to nonflat backgrounds with torsion, corresponding to the Einstein-Cartan-Kibble-Sciama theory treating the translational and rotational symmetries separately. By varying with respect to the tetrads and connections they obtained the expressions \cite{butcher2} which in the harmonic gauge reduce to the results found in \cite{butcher}, \cite{butcher3}. In this sense the expressions follow as Noether currents associated with the symmetries under translations and rotations.

The main motivations for the present paper have been to find (i) relations between our past work \cite{bicak}, \cite{bicak2} to the Landau-Lifshitz complex employed with an increasing frequency in the literature (cf. e.g. the recent monograph \cite{will}), (ii) to give relations to new developments due to the Cambridge group \cite{butcher}, \cite{butcher2} and (iii) to generalize our method of studying the uniqueness of energy-momentum tensors to the case of massive gravity. The massive gravity has been studied with an ``oscillatory interest'' for the past 70 years. It became popular again recently when it was proven that the nonlinear theory of massive gravity is ghost free \cite{hassan1}, \cite{hassan2}; see also the reviews \cite{bichler}, \cite{derham}. Here, we shall consider just the case of the linear Fierz-Pauli theory; it represents the weak field limit of generic theories. As far as we are aware, no study of the uniqueness of the energy-momentum tensor in massive gravity was done so far. Last but not least, we wish to demonstrate how long and tedious calculations which were necessary to get results in \cite{bicak}, \cite{bicak2}, and, also, how other procedures like finding appropriate superpotentials can, at present, be performed very effectively by the usage of symbolic tensor manipulation software \textsc{Cadabra}.

The article is organized as follows. In the following Sec. \ref{secgenmethod} we describe a general procedure of finding energy-momentum tensors conserved as a consequence of a system of equations of motion given by a system of partial differential equations containing linearly second derivatives of the second-rank tensor $h_{ik}$.

The important step in simplifying computations is to consider, in Sec. \ref{seclorentz}, Lorentz covariant expressions; this does not mean any loss of generality. We construct a general second-rank tensor quadratic in $h_{ik,l}$; it involves 20 free constant parameters. We discuss the conservation of the tensor as a consequence of the field equations of various types. It is here where the use of the \textsc{Cadabra} software is indicated. More details are relegated to the Appendix.

In Sec. \ref{massiveprocedure}, the method is generalized to the equations of motion containing nonderivative terms $h_{ik}$ which is the case of the Fierz-Pauli theory of massive gravity.

It is well known that there may exist parts of energy-momentum tensors which do not contribute to the total quantities for insular systems with fields decaying sufficiently rapidly at infinity. These ``generalized'' divergences called superpotentials can be investigated again by the modification of the method presented in the previous sections. In Sec. \ref{secsuperpot} a general expression with 13 arbitrary parameters is constructed and the form of the master equation for the superpotential is given. It combines the condition that the divergence of the superpotential must yield the energy-momentum tensor as a consequence of the field equations.

Sections \ref{secresultlingrav} and \ref{massivegravitychapter} contain the results. Conserved quantities in linearized gravity are discussed in Sec. \ref{secresultlingrav}. First, a unique albeit nonsymmetric expression (and corresponding superpotential) without using equations of motion is presented in Sec. \ref{tstrong}. It appears in a number of subsequent expressions. Next, the linearized vacuum Einstein's equations are employed in Sec. \ref{tbicak} to obtain a four-parametric family of energy-momentum tensors quadratic in $h_{ik,l}$. The condition of symmetry leads to the \textit{unique} (up to a multiplicative constant) result which is just the linearized Landau-Lifshitz pseudotensor. We give also ``the metric energy-momentum tensor'' (so automatically symmetric) which follows from the variational principle and is covariantly conserved at any background. However, it contains the second derivatives $h_{ik,lm}$. In this first part of Sec. \ref{secresultlingrav} we proceed and find quantities conserved in a completely general gauge.

Within the linearized gravity we also investigate the role of the harmonic gauge condition and generalized gauge condition since we wish to analyze the uniqueness of the energy-momentum tensor presented in \cite{butcher}. Taking into account the harmonic gauge condition \textit{ab initio}, our procedure gives the five-parameter family of, generally nonsymmetric, conserved quantities. The condition of symmetry leaves us with a three-parameter expression which contains the resulting tensor given in \cite{butcher} as a special case. However, starting from the unique symmetric energy-momentum tensor obtained without any gauge condition and applying then the harmonic gauge condition \textit{a posteriori}, we do \textit{not} arrive at the result in \cite{butcher}. In the most recent work \cite{butcher2}, the authors rederive their symmetric tensor from a variational principle without a using special gauge condition -- they apply the harmonic gauge condition \textit{a posteriori}. The ``initial'' tensor obtained in \cite{butcher2} is nonsymmetric; it follows as a special case from our four-parameter family of energy-momentum tensors.

In \cite{butcher} the authors start with a generalized gauge condition, $h{_{ab}}^{,b} = \chi h_{,a}$, where $h = \eta^{ik} h_{ik} = h{_{i}}^i$, and the harmonic gauge condition is found to be a consequence of their physical arguments leading to $\chi = \frac{1}{2}$. Using our procedure we also construct conserved expressions for arbitrary values of $\chi$. The case $\chi = 1$ leads to the four-parameter family and $\chi \neq 1$ produces the five-parameter family. The requirement that the resulting energy-momentum tensor is conserved independently of $\chi$ leads to \textit{a unique nonsymmetric} expression.

Energy-momentum tensors for massive gravity stemming from the equations following from the Fierz-Pauli action are constructed in Sec. \ref{massivegravitychapter}. Starting first with the Klein-Gordon equation of the form $\Box h_{ik} - m^2 h_{ik} = 0$ in Sec. \ref{subKGeqn}, we obtain a five-parameter family of conserved expressions. If we add additional equations of the Fierz-Pauli gravity, $h{_{ab}}^{,b} = 0$, $h=0$, the system of conserved energy-momentum tensors reduces to the three-parameter family, and the condition of symmetry yields a two-parameter family. However, we can arrive at a unique expression in the following way. Rather than from the Klein-Gordon equation for massive field $h_{ik}$, we start from the field equation as it follows directly from the Fierz-Pauli action. The resulting tensors are nonsymmetric and form a two-parameter system. Nevertheless, after inserting conditions $h{_{ab}}^{,b} = 0$, $h=0$, and imposing the symmetry of the energy-momentum tensor we arrive at the unique expression. It is different from the generalization of the linearized Landau-Lifshitz tensor to the case of massive gravity but it yields the same total quantities since both expressions differ by the divergence of a superpotential. It also differs by the divergence of a superpotential from the canonical energy-momentum tensor derived from the variational principle based on the Fierz-Pauli action.

\section{The method of finding a general conserved energy-momentum complex \label{secgenmethod}}

We wish to construct a conserved energy-momentum complex\footnote{In full general relativity one cannot form a true energy-momentum tensor -- various nontensorial objects suggested are called ``complexes'' or ``pseudotensors'' (\cite{MTW}, \cite{landau}). In their linearized versions, however, they transform as tensors under Lorentz transformations though they are not invariant under the gauge transformation $x^i \rightarrow x^{i'} = x^i + \xi^i$.} $T^{ij}$ for the linearized gravity which depends quadratically on the first derivatives of the metric. So we assume its form to be
\begin{equation}
  T^{ij} = t^{ijabcrst} h_{ab,c} h_{rs,t},
\end{equation}
with $t^{ijabcrst}$ being constant coefficients symmetric in $(a,b)$ and $(r,s)$ and invariant with respect to the interchange of the triples $(a,b,c)$ and $(r,s,t)$. In vacuum it has to satisfy the conservation law
\begin{equation}
  T^{ij}_{,j} = 0
\end{equation}
as a consequence of the equations of motion assumed, just here, to be in the form
\begin{equation}
  P^{A} = p^{Amnop} h_{mn,op} = 0; \label{moteqncoeff}
\end{equation}
$A$ is an arbitrary multi-index; $p$'s are constant coefficients. Using Lagrange multipliers $\lambda_A^i$ these requirements can be written as the following master equation 
\begin{equation}
  T^{ij}_{,j} = \lambda^i_A P^A \label{mastereqn}
\end{equation}
which is assumed to be satisfied for arbitrary independent field variables; hence the divergence of the energy-momentum tensor is formed from a linear combination of the field equations. Lagrange multipliers $\lambda^i_A(x)$ are in general functions of spacetime coordinates. Since $T^{ij}_{,j} = 2 t^{ijabcrst} h_{ab,c} h_{rs,tj}$, the Lagrange multipliers in this case need to have the form
\begin{equation}
  \lambda^i_A = L^{iabc}_A h_{ab,c}, 
\end{equation}
where $L$'s are constant coefficients. Writing master equation (\ref{mastereqn}) in terms of coefficients $t^{\ldots}$ and $L^{\ldots}_A$ we have
\begin{equation}
 ( 2 t^{ipabcmno} - L^{iabc}_A p^{Amnop} ) h_{ab,c} h_{mn,op} = 0. \label{koefextrakt}
\end{equation}
The last equation has to be satisfied for all $h_{ab}$ and their derivatives.  Taking into account the obvious symmetries we arrive at the condition
\begin{equation}
  t^{ip(ab)c(mn)o} + t^{io(ab)c(mn)p} - L^{i(ab)c}_A p^{A(mn)(op)} = 0. \label{symcoeff}	
\end{equation}
Here $()$ denotes symmetrization, $[]$ used below -- antisymmetrization, both with $\frac{1}{2}$ included. The final step in this general method consists of eliminating Lagrange multipliers $L^{iabc}_A$ employing known coefficients $p^{Amnop}$ and so find the constants $t^{ijabcrst}$.

\section{Lorentz covariant theories \label{seclorentz}}

Assuming that field equations and the corresponding energy-momentum tensor are Lorentz covariant the procedure described above considerably simplifies. Raising and lowering indices will be performed by the Minkowski metric $\eta_{ab}$. Now we just need to find all different contractions of the term $h_{ab,c} h_{rs,t}$ to produce a tensor of rank two. The most general form of a Lorentz covariant energy-momentum tensor quadratic in the first derivatives of the metric then turns out to contain 20 parameters $a_1$,..., $a_{20}$. It reads as follows:
\begin{align}
  T_{ik} = \,\, &  a_1  h_{ik,a} h{^{ab}}_{,b} + a_2 h_{ik,a} h^{,a} + a_3 h{_{ia}}^{,a} h{_{kb}}^{,b}  \nonumber \\ 
  & + a_4 h_{ia,b} h{_k}^{a,b} + a_5 h_{ia,b} h{_k}^{b,a} + a_6 h_{ia,k} h{^{ab}}_{,b}  \nonumber \\
  & + a_7 h_{ka,i} h{^{ab}}_{,b} + a_8 h_{ia,k} h^{,a} + a_9 h_{ka,i} h^{,a}  \nonumber \\ 
  & + a_{10} h{_{ia}}^{,a} h_{,k} + a_{11} h{_{ka}}^{,a} h_{,i} + a_{12} h_{ia,b} h{^{ab}}_{,k}   \nonumber \\
  & + a_{13} h_{ka,b} h{^{ab}}_{,i} + a_{14} h_{,i} h_{,k} + a_{15} h_{ab,i} h{^{ab}}_{,k}   \nonumber \\ 
  & + a_{16} \eta_{ik} h_{,b} h{^{bc}}_{,c} + a_{17} \eta_{ik} h{_{ab}}^{,a} h{^{bc}}_{,c} + a_{18} \eta_{ik} h_{,b} h^{,b}   \nonumber \\
  & + a_{19} \eta_{ik} h_{ab,c} h^{ab,c} + a_{20} \eta_{ik} h_{ab,c} h^{bc,a}. \label{SET}
\end{align}

In order to simplify the notation of some expressions in the following we shall denote a term appearing at a particular coefficient $a_{\alpha}$ by $\mathcal{A}_{\alpha \, ik}$ ($\alpha=1$,...,\,$20$). The energy-momentum tensor and its divergence can thus be written as
\begin{equation}
  T_{ik} = \sum_{\alpha=1}^{20} a_{\alpha} \mathcal{A}_{\alpha \, ik}, \qquad T{_{ik,}}^k = \sum_{\alpha=1}^{20} a_{\alpha} \mathcal{A}{_{\alpha \, ik}}^{,k}.
\end{equation}

Let us now consider various types of equations of motion, in a ``tensorial form,'' depending on the number of their free indices, $P_{ab} = 0$
(e.g., the case of the linearized Einstein equations), $P_a = 0$ (e.g., the equations characterizing gauge conditions or field equations in the case of massive gravity), and $P = 0$ (the case of massive gravity).

In the first case we assume that $P_{ab} = P_{ba}$ contain linearly $h_{mn,op}$. Regarding our ansatz for energy-momentum tensor (\ref{SET}) the right-hand of the master equation (\ref{mastereqn}) acquires the form $\lambda_i^{rsqab} h_{rs,q} P_{ab}$. After taking into account the Lorentz covariance and considering all relevant symmetries we find, explicitly, the resulting contribution to the master equation:
\begin{align}
	\sum_{\beta=1}^6 \lambda_{\beta} \mathcal{L}_{\beta i} = \,\,& \lambda_1 \, h^{,a} P_{ia} + \lambda_2 \, h{^{ab}}_{,b} P_{ia} + \lambda_3 \, h{_{ib}}^{,b} P_a^a  \nonumber \\
	& + \lambda_4 \, h_{,i} P_a^a + \lambda_5 \, h{_i}^{a,b} P_{ab} + \lambda_6 \, h{^{ab}}_{,i} P_{ab}; \label{lambdaprispevek}
\end{align}
here $\lambda_{\beta}$ are scalar Lagrange multipliers and $\mathcal{L}_{\alpha \, i}$ denote corresponding terms.

Analogously, we proceed in the case of the field equation with the vectorial form $P_a = 0$. For our purposes we consider the field equations linear in $h_{ab,c}$. Therefore, in the master equation there will appear the term $\mu_i^{mnopa} h_{mn,op} P_a$ with the explicit form
\begin{align}
	\sum_{\beta=1}^6 \mu_{\beta} \, \mathcal{U}_{\beta \, i} = \,\,& \mu_1 \, h{_{,b}}^b P_i + \mu_2 \, h{_{ab}}^{,ab} P_i + \mu_3 \, h{_{ia}}^{,ab} P_b \nonumber \\
	& + \mu_4 \, h{_{ia,b}}^b P^a + \mu_5 \, h{_{,i}}^b P_b + \mu_6 \, h{_{ab,i}}^a P^b,
\end{align}
where $\mu_{\beta}$ are scalar Lagrange multipliers and the individual terms are labeled as $\mathcal{U}_{\alpha \, i}$.

Finally, consider the equation $P=0$. Our linearity condition and the general form of the energy-momentum tensor restrict the possible choice just to $P = h{{{_a}^a}_{,b}}^b$. Nevertheless, in the master equation there will arise the term $ \kappa_i^{qrs} h_{rs,q} P$ leading to two covariant terms called $\mathcal{K}_{\alpha \, i}$, with Lagrange multipliers $\kappa_{\alpha}$:
\begin{equation}
	\sum_{\beta=1}^2 \kappa_{\beta} \mathcal{K}_{\beta\,i} = \kappa_1 \, h{_{ia}}^{,a} P + \kappa_2 \, h_{,i} P.
\end{equation}

Summarizing the previous considerations, we find the master equation in the following general form
\begin{equation}
	\sum_{\alpha=1}^{20} a_{\alpha} \mathcal{A}{_{\alpha \, ik}}^{,k} = \sum_{\beta=1}^6 \lambda_{\beta} \mathcal{L}_{\beta \, i} + \sum_{\beta=1}^6 \mu_{\beta} \mathcal{U}_{\beta \, i} +  \sum_{\beta=1}^2 \kappa_{\beta} \mathcal{K}_{\beta \, i}. \label{covariantmaster}
\end{equation}
As a result we obtain equations for unknowns $a_{\alpha}$, $\lambda_{\beta}$, $\mu_{\beta}$, and $\kappa_{\beta}$ which have to hold for arbitrary field variables $h_{ij}$. We rewrite them in the form of general equation (\ref{koefextrakt}), though Lorentz covariance substantially reduces the number of terms. As a consequence of the linear independence of the terms $h_{ab,c} h_{mn,op}$, we can extract linear equations for variables $a_{\alpha}$, $\lambda_{\beta}$, $\mu_{\beta}$, and $\kappa_{\beta}$. This extraction can be assisted by the use of the \textsc{Cadabra} software. We illustrate its use in our context in the Appendix.

\section{The case of massive gravity \label{massiveprocedure}}

Above, we considered the equations of motion containing linearly $h_{ab,c}$ or $h_{mn,op}$. We now generalize the procedure to allow field $h_{ab}$ itself to be present linearly in equations of motion as, for example, in the Klein-Gordon-type equation $h{_{ab,c}}^c - m^2 h_{ab} = 0$, or in the Fierz-Pauli equation $h{_{ab,c}}^c - h{_{ac,b}}^c - h{_{bc,a}}^c + \ldots - m^2 \left( h_{ab} - \eta_{ab} h \right) = 0$ which we shall consider in detail in Sec. \ref{massivegravitychapter}.

In this more general case we assume the energy-momentum tensor to contain not only quadratic terms in the first derivatives of the metric but also the terms of the form $h_{ab} h_{cd}$ appropriately contracted to give a tensor of rank two.\footnote{Notice that the terms of the form $h_{ab} h_{cd,e}$ will not yield a tensor of rank two.} There are just four terms of this type
\begin{equation}
	\sum_{\beta=1}^4 c_{\beta} \, \mathcal{C}_{\beta ik} = c_1 \, h_{ik} h + c_2 \, h_{ia} h_k^a + c_3 \, \eta_{ik} h^2 + c_4 \, \eta_{ik} h_{ab} h^{ab}, \label{additionalSET}
\end{equation}
where $\mathcal{C}_{\beta \, ik}$ just denote terms explicitly seen on the right-hand side.\footnote{We did not consider these $\mathcal{C}_{\beta}$-terms in the previous section since they would vanish anyway, because the equations of motion involve only the second derivatives and whatever choice of multipliers $\lambda^{\cdots}_A$ will not produce the terms $h_{ab} h_{cd,e}$ occuring in $T^{ik}_{,k}$.} Therefore, the general form of the energy-momentum tensor we consider, in the case of massive gravity, for example, will read as follows:
\begin{equation}
	T_{ik} = \sum_{\alpha=1}^{20} a_{\alpha} \mathcal{A}_{\alpha \, ik} + \sum_{\beta=1}^4 c_{\beta} \mathcal{C}_{\beta \, ik}. \label{generalizedSET}
\end{equation}

Considering next the equation of motion we have now to modify relation (\ref{moteqncoeff}) into $P_{ab} = p{_{ab}}^{mnop} h_{mn,op} + q{_{ab}}^{mn} h_{mn} = 0$. The character of equations of motion assumed and our ansatz for the energy-momentum tensor imply that the Lagrange multipliers are linear in the first derivatives of $h_{ab}$, $\lambda^i_{ab} = \lambda^{irst}_{ab} h_{rs,t}$.

In the case of the vector-type field equations, $P_a = 0$, we now get an additional contribution to the master equation, $\nu_i^{mna} h_{mn} P_a$, which leads to two covariant terms labeled by $\mathcal{V}_{\alpha}$, with Lagrange multipliers $\nu_{\alpha}$:
\begin{equation}
	\sum_{\beta=1}^2 \nu_{\beta} \mathcal{V}_{\beta} = \nu_1 \, h{_a}^a P_i + \nu_2 \, h{_i}^a P_a. 
\end{equation}

For the scalar-type field equation, $P = 0$, a new term $P = h_a^a$ can arise. It will appear in Sec. \ref{subKGeqn}.

\section{Superpotentials \label{secsuperpot}}

It is of interest to know whether some part of an energy-momentum tensor can be derived from a so-called superpotential. Under suitable boundary conditions this part does not contribute to total quantities. We now describe the general method  of constructing superpotentials, later we use it in specific cases. The energy-momentum tensor $T_{ik}$ is generated by the superpotential $U_{ikl} = U_{i[kl]}$ if the following master equation holds
\begin{equation}
	U{_{ikl,}}^{l} = T_{ik} + \lambda{_{ik}}^{A} P_{A};
\end{equation}
i.e. the divergence of a superpotential gives the given energy-momentum tensor and a linear combination of field equations $P_{A} = 0$ with multipliers $\lambda{_{ik}}^A$. The antisymmetry in indices $(k,l)$ then implies the conservation law $T{_{ik}}^{,k} = U{_{i[kl]}}^{,(kl)} = 0$.
	The terms $h_{ab} h_{cd}$ present in the case of massive gravity cannot be produced by a divergence; hence we will restrict our attention to tensors $T_{ik}$ quadratic in the first derivatives of the metric, $h_{ab,c} h_{de,f}$ -- these can be produced by the divergence of terms of the form $h_{ab} h_{cd,e}$.

The requirement of the Lorentz covariance, the antisymmetry, and the structure of superpotential $U_{ikl} \propto h_{ab} h_{cd,e}$ lead to a general expression with 13 parameters as follows:
\begin{align}
  U_{ikl} = \,\,& \sum_{\alpha = 1}^{13} u_{\alpha} \, \mathcal{U}_{\alpha \, ikl} = u_1 h_{i[k} h{_{l]a}}^{,a} + u_2 h_{i[k} h_{,l]} \nonumber \\
  & + u_3 h_{ia} h{^{a}}_{[k,l]} + u_4 h_{a[k} h{_{l]i}}^{,a} + u_5 h{^{a}}_{i,[k} h_{l]a}  \nonumber \\
  & + u_6 h_{a[k} h{^{a}}_{l],i} + u_7 h \, h_{i[k,l]} + u_8 \eta_{i[k} h_{l]a} h{^{ab}}_{,b}  \nonumber \\
  & + u_9 \eta_{i[k} h_{l]a} h^{,a} + u_{10} \eta_{i[k} h \, h{_{l]a}}^{,a} + u_{11} \eta_{i[k} h^{ab} h_{l]a,b}  \nonumber \\
  & + u_{12} \eta_{i[k} h \, h_{,l]} + u_{13} h_{ab} h{^{ab}}_{,[k} \eta_{l]i}. \label{suppot}
\end{align}
Considering the equations of motion with two indices, $P_{ab}$, which contain linearly the second derivatives of field variables $h_{ab}$, the Lagrange multipliers $\lambda{_{ik}}^{ab}$ will be proportional just to $h_{ab}$. The resulting Lorentz covariant expression for $\lambda{_{ik}}^{ab} P_{ab}$ is
\begin{align}
	\lambda{_{ik}}^{ab} P_{ab} = \,\,& \lambda_1 h_{ik} P + \lambda_2 h_{ia} P{_{k}}^{a} + \lambda_3 h_{ka} P{_{i}}^{a} \nonumber \\
	& + \lambda_4 h P_{ik} + \lambda_5 \eta_{ik} h \, P{_{a}}^a.
\end{align}

In practice we are solving just the equations involving the second derivatives $h_{ab,cd}$, i.e.
\begin{equation}
	\left( U{_{ikl}}^{,l} \right)_{\mbox{\tiny 2nd derivatives}} = \lambda{_{ik}}^{ab} P_{ab}.
\end{equation}
This restricts the coefficients $u_{\alpha}$ in the general expression (\ref{suppot}). The resulting superpotential-generated tensors $T_{ik}$ are then easily computed as $T_{ik} = U{_{ikl}}^{,l}$.

\section{Conserved quantities in the linearized gravity \label{secresultlingrav}}

In the first part of this section we find the second-rank tensors constructed from the quadratic expressions in $h_{ik,l}$ and conserved as a consequence of the linearized Einstein equations \textit{without choosing any particular gauge}. In the second part (Secs. \ref{chgauge}, \ref{chgaugeparam}) we first impose the harmonic and generalized harmonic gauges and look for the expressions conserved under these conditions. In this way we find, among others, under which conditions we arrive at the results obtained by Butcher \textit{et al.} \cite{butcher}, \cite{butcher2}.

It is well known that, in contrast to the linearized curvature tensor, quantities involving the first derivatives $h_{ik,l}$ are gauge dependent. At the end of Sec. \ref{secresultlingrav} we note that in the high-frequency case, after suitable averaging introduced by Isaacson \cite{isaacson1}, \cite{isaacson2}, the expressions become gauge invariant and can be calculated for all choices of gauge.

\subsection{Strongly conserved quantity \label{tstrong}}

Let us first consider a possibility whether there exists a combination of parameters $a_i$ for which the tensor (\ref{SET}) is conserved identically, i.e., without using field equations. It turns out that, indeed, such a tensor exists for the choice of constants $a_i$ vanishing except for $a_7 = - a_{13} = -2 a_{17} = 2 a_{20}$. Denoting this one free parameter by $\alpha \, ( \, = a_7 = \ldots \, )$, the resulting strongly conserved quantity is uniquely given by
\begin{align}
  T_{ik} = \,\,& \alpha \left( h_{ka,i} h{^{ab}}_{,b} - h_{ka,b} h{^{ab}}_{,i} - \frac{1}{2} \eta_{ik} h{_{ab}}^{,a} h{^{bc}}_{,c} \right. \nonumber \\
  & \left. \quad + \frac{1}{2} \eta_{ik} h_{ab,c} h^{bc,a} \right) = \alpha \, T_{ik}^{(strong)},
\end{align}
where, for future reference, we denoted the expression in brackets as $T_{ik}^{(strong)}$.

This conserved tensor is generated by the superpotential
\begin{equation}
	U_{ikl} = \alpha \left( h_{a[k} h{^{a}}_{l],i} + \eta_{i[k} h_{l]a} h{^{ab}}_{,b} - \eta_{i[k} h^{ab} h_{l]a,b}  \right). \label{strongsuperpotential}
\end{equation}

\subsection{Linearized vacuum Einstein's equations \label{tbicak}}

We now allow the divergence of the energy-momentum tensor to be a general linear combination of the linearized Einstein's field equations: $T^{ik}_{,k} = \lambda^{irs} R_{rs}$, where $R_{rs}$ is the linearized Ricci tensor. The resulting tensor depends on four free parameters which we denote $\alpha_1,\alpha_2,\alpha_3,\alpha_4$. The relation between the constant parameters $a_i$ from (\ref{SET}) and parameters $\alpha_i$ is $\alpha_1 = a_9 = a_{15} = - 2 a_{19}$, $\alpha_2 = a_7 = - 2 a_{17}$, $\alpha_3 = - a_{11} = a_{14} = a_{16}$, $\alpha_4 = a_1 = - a_3 = a_4 = a_{10} = - a_{12}$, $a_5 = a_6 = 0$, $\alpha_1 - \alpha_3 = - a_8$, $2 \alpha_1 + \alpha_2 = - a_{13} = 2 a_{20}$, $\frac{1}{2} \alpha_1 + \alpha_3 = - a_{16}$, $\alpha_1 + \alpha_3 - \alpha_4 = a_2$. The final form of a general tensor conserved as a consequence of vacuum equations of linear gravity thus looks as follows:
\begin{align}
  T_{ik} = \,\, & \alpha_1 \Big( h_{ik,a} h^{,a} - h_{ia,k} h^{,a} + h_{ka,i} h^{,a} - 2 h_{ka,b} h{^{ab}}_{,i}   \nonumber \\
  & \qquad + h_{ab,i} h{^{ab}}_{,k} - \frac{1}{2} \eta_{ik} h_{,b} h^{,b} - \frac{1}{2} \eta_{ik} h_{ab,c} h^{ab,c}   \nonumber \\
  & \qquad + \eta_{ik} h_{ab,c} h^{bc,a}  \Big) + \alpha_2 \, T_{ik}^{(strong)} \, + \nonumber \\
		& \, \alpha_3 \big( h_{ik,a} h^{,a} - h_{ia,k} h^{,a} - h{_{ka,}}^a h_{,i} + h_{,i} h_{,k}   \nonumber \\
	& \qquad \left. + \, \eta_{ik} h_{,b} h{^{bc}}_{,c} - \eta_{ik} h_{,b} h^{,b}  \right) + \nonumber \\
  		  & \, \alpha_4 \left( h_{ik,a} h{^{ab}}_{,b} - h_{ik,a} h^{,a} - h{_{ia}}^{,a} h{_{kb}}^{,b}  \right. \nonumber \\
  	& \qquad \left. + \, h_{ia,b} h{_k}^{a,b} + h{_{ia}}^{,a} h_{,k} - h_{ia,b} h{^{ab}}_{,k} \right). \label{gauge1general}
\end{align}

It involves a four-parameter freedom; one of the parameters can be fixed by the choice of units. A natural question arises whether among these expressions there exist quantities which are symmetric, $T_{ik} = T_{ki}$; this condition imposes some restrictions on coefficients $\alpha_i$. We obtain \textit{a unique} (up to a multiplicative constant $\alpha$) symmetric tensor writing $\alpha_1 = \alpha$, $\alpha_2 = 0$, $\alpha_3 = - 2\alpha$, $\alpha_4 = 2 \alpha$. Putting $\alpha_i$'s into (\ref{gauge1general}) we arrive at the final expression in the form
\begin{align}
  T_{ik}^{(sym)} = \,\, & \alpha \, \bigg( 2 h_{ik,a} h{^{ab}}_{,b} - 3 h_{ik,a} h^{,a} - 2 h{_{ia}}^{,a} h{_{kb}}^{,b}   \nonumber \\
  & \qquad + 2 h_{ia,b} h{_k}^{a,b} + 2 h_{a(i,k)} h^{,a} + 4 h_{,(i} h{_{k)a}}^{,a}  \nonumber \\
  & \qquad - 4 h_{ab,(i} h{_{k)}}^{a,b} - 2 h_{,i} h_{,k} + h_{ab,i} h{^{ab}}_{,k}  \nonumber \\
  & \qquad - 2 \eta_{ik} h_{,b} h{^{bc}}_{,c} + \frac{3}{2} \eta_{ik} h_{,b} h^{,b}  \nonumber \\
  & \qquad \left. - \frac{1}{2} \eta_{ik} h_{ab,c} h^{ab,c} + \eta_{ik} h_{ab,c} h^{bc,a} \right). \label{symetrickybicak}
\end{align}

The four-parameter family of conserved quantities in the linearized theory was in fact obtained in \cite{bicak} already without using \textsc{Cadabra}; however individual terms given there contain a number of misprints\footnote{Denoting the parameters $\stackrel{i}{\alpha}$ used in \cite{bicak} by $\beta_i$ we obtain the relationships between parametrization used above and in \cite{bicak}: $\alpha_1 = 2 \beta_3$, $\alpha_2 = - 2 \beta_4$, $\alpha_3 = \beta_1$, $\alpha_4 = - \beta_2$. Condition of symmetry yields $\beta_1 = 2 \beta$, $\beta_2 = 2 \beta$, $\beta_3 = -\frac{1}{2} \beta$, $\beta_4 = 0$.}.

Notice that we could also start out from the ``complete'' linearized Einstein's equations $G_{rs} = 0$, where $G_{rs}$ is the Einstein tensor, and consider the master equation $T^{ik}_{,k} = \lambda^{irs} G_{rs}$. The result, as expected, will not change; it only leads to regular linear transformations of Lagrange multipliers because of the following identity
\begin{align}
  \lambda^{irs} G_{rs} &= \lambda'^{\,icd} R_{cd}, & \lambda'^{\,icd} &= \lambda^{irs} \left( \delta^c_r \delta^d_s - \frac{1}{2} \eta_{rs} \eta^{cd} \right).
\end{align}

If we look for superpotentials generating parts of the conserved energy-momentum tensor in the linearized gravity, we find that expressions multiplied by $\alpha_2$, $\alpha_3$, and $\alpha_4$ can be expressed as a divergence of the following expression:
\begin{align}
  U_{ikl} = \,\, & \alpha_2 \left( h_{a[k} h{^{a}}_{l],i} + \eta_{i[k} h_{l]a} h{^{ab}}_{,b} - \eta_{i[k} h^{ab} h_{l]a,b} \right) + \nonumber \\
  & 2 \alpha_3 \left( h \, h_{i[k,l]} + \eta_{i[k} h \, h{_{l]a}}^{,a} - \eta_{i[k} h \, h_{,l]} \right) + \nonumber \\
  & 2 \alpha_4 \left( h_{i[k} h{_{l]a}}^{,a} - h_{i[k} h_{,l]} + h_{ia} h{^{a}}_{[k,l]} \right). \label{superbicak}
\end{align}
Therefore, the general energy-momentum tensor (\ref{gauge1general}) can be written in the form
\begin{align}
	T_{ik} = \,\, & \alpha_1 \bigg( h_{ik,a} h^{,a} - h_{ia,k} h^{,a} + h_{ka,i} h^{,a} - 2 h_{ka,b} h{^{ab}}_{,i} \nonumber \\
	& \qquad + h_{ab,i} h{^{ab}}_{,k} - \frac{1}{2} \eta_{ik} h_{,b} h^{,b} - \frac{1}{2} \eta_{ik} h_{ab,c} h^{ab,c} \nonumber \\
	 & \qquad + \eta_{ik} h_{ab,c} h^{bc,a} \bigg) + U{_{ikl}}^{,l}, \label{gauge1generalsuperpot}
\end{align}
where $U_{ikl}$ is given by (\ref{superbicak}), whereas the symmetric tensor (\ref{symetrickybicak}) can be written as
\begin{align}
	T_{ik}^{(sym)} = \,\, & \alpha \bigg[ h_{ik,a} h^{,a} - h_{ia,k} h^{,a} + h_{ka,i} h^{,a} \nonumber \\ 
	& \quad - 2 h_{ka,b} h{^{ab}}_{,i} + h_{ab,i} h{^{ab}}_{,k} - \frac{1}{2} \eta_{ik} h_{,b} h^{,b}  \nonumber \\
	& \quad - \frac{1}{2} \eta_{ik} h_{ab,c} h^{ab,c} + \eta_{ik} h_{ab,c} h^{bc,a}  \nonumber \\
	& \quad + 4 \left( h \, h_{i[k,l]} + \eta_{i[k} h \, h{_{l]a}}^{,a} - \eta_{i[k} h \, h_{,l]}  \right. \nonumber \\
	& \quad \left. - \, h_{i[k} h{_{l]a}}^{,a} + h_{i[k} h_{,l]} - h_{ia} h{^{a}}_{[k,l]} \right)^{,l} \bigg], \label{gauge1generalsuperpotsym}
\end{align}
in which the terms in the round brackets form a superpotential.

\subsubsection{Energy-momentum tensor obtained by variational principle \label{subsubvari}}

It is worth to mention the result indicated in the text of a lecture in \cite{bicak2}. We start from the covariant Lagrangian density for the tensor field $h_{ab}$ representing linear perturbations of the vacuum background spacetime metric $g_{ab}$. It has the form
\begin{equation}
	(-g)^{-\frac{1}{2}} L = \frac{1}{2} h_{ab;c} h^{ab;c} - \frac{1}{2} h_{;a} h^{;a} + h_{;a} h{^{ab}}_{;b} - h_{ab;c} h^{bc;a},
\end{equation}
where covariant derivatives are done with respect to the background metric $g_{ab}$. The metric energy-momentum tensor following from the variational principle reads
\begin{align}
	T_{ik} = \,\,& \frac{2}{\sqrt{-g}} \frac{\delta L}{\delta g^{ik}} =  g_{ik} \left( \frac{1}{2} h_{ab;c} h^{ab;c} - \frac{1}{2} h_{;a} h^{;a} \right. \nonumber \\
	& - h_{ab;c} h^{bc;a} - h_{;ab} h^{ab} \bigg) - h_{ab;i} h{^{ab}}_{;k} + h_{;i} h_{;k}  \nonumber \\
	& - 2 h_{;(i} h{_{k)a}}^{;a} + 4 h_{ab;(i} h{_{k)}}^{a;b} - 2 h_{ia;b} h{_k}^{a;b} \nonumber \\
	& - 2 h_{ia;b} h{_k}^{b;a} + 2 h_{ik;a} h{^{ab}}_{;b} + h_{ik;a} h^{;a}  \nonumber \\
	& + 2 h_{;a(i} h^{a}_{k)} - 4 h_{a(i} h{_{k)b}}^{;ab} + 2 h_{ik;ab} h^{ab} + h_{ik} h{_{;a}}^a. \label{subsubvaritensor}
\end{align}
The resulting energy-momentum tensor contains \textit{the second derivatives} of field $h_{ab}$ and, even with flat background, cannot thus be obtained by our procedure. However, it is worthwhile to notice that it is covariantly conserved in a general background spacetime.

\subsubsection{Linearized Landau-Lifshitz pseudotensor \label{subsublandau}}

Consider the Landau-Lifshitz energy-momentum pseudotensor in the full general relativity (see e.g.  \cite{MTW}, \cite{landau}, \cite{will})
\begin{align}
	& 16 \pi \, (-g) \, t^{ab} = \hat{g}{^{ab}}_{,c} \, \hat{g}{^{cd}}_{,d} - \hat{g}{^{ac}}_{,c} \, \hat{g}{^{bd}}_{,d} + \frac{1}{2} g^{ab} g_{cd} \, \hat{g}{^{ce}}_{,f} \, \hat{g}{^{fd}}_{,e} \nonumber \\
	& \qquad - g_{cd} \,  \hat{g}{^{ce}}_{,f} \left( g^{af} \hat{g}{^{bd}}_{,e} + g^{bf} \hat{g}{^{ad}}_{,e} \right) \nonumber \\
		& \qquad + g_{cd} g^{ef} \, \hat{g}{^{ac}}_{,e} \hat{g}{^{bd}}_{,f} + \frac{1}{8} \left( 2g^{ac} g^{bd} - g^{ab} g^{cd} \right) \cdot \nonumber \\
		& \qquad  \cdot \Big( 2 g_{ef} g_{mn} - g_{fm} g_{en} \Big) \hat{g}{^{en}}_{,c} \, \hat{g}{^{fm}}_{,d}, \label{LLtensor} 
\end{align}
where $g_{ab}$ is a spacetime metric and $\hat{g}^{ab}$ denotes $\sqrt{-g} \, g^{ab}$; $g = \det(g_{ab})$. If we now use the linearization ansatz $g_{ab} = \eta_{ab} + h_{ab}$, $g^{ab} = \eta^{ab} - h^{ab}$, where $h^{ab} = \eta^{ac} \eta^{bd} h_{cd}$, we find that $\hat{g}{^{ab}}_{,c} = \frac{1}{2} \eta^{ab} h_{,c} - h{^{ab}}_{,c} + O(h^2)$. Writing out the terms up to the second order in Landau-Lifshitz pseudotensor (\ref{LLtensor}), which is tedious but straightforward, \textit{we get the symmetric energy-momentum tensor} (\ref{symetrickybicak}).

\subsubsection{Harmonic gauge condition \label{chgauge}}

We now wish to analyze the uniqueness of the energy-momentum tensor suggested recently in \cite{butcher}. Since there the assumption of the linearized harmonic gauge condition 
\begin{equation}
	h^{ab}_{,b} = \frac{1}{2} h^{,a} \label{harmonicgauge}
\end{equation}
plays a fundamental role, we have to generalize the previous procedure to include this possibility. A similar condition will become the field equation in the case of massive gravity considered in Sec. \ref{massivegravitychapter}.

We could just add the gauge condition and its derivatives multiplied by another set of Lagrange multipliers. However, with this simple gauge condition our procedure is equivalent to the following. First, regarding the gauge condition (\ref{harmonicgauge}), we replace all terms $h^{ab}_{,b}$ appearing in general expression (\ref{SET}) by $\frac{1}{2} h^{,a}$. Then, we observe that some terms in (\ref{SET}) will become equal: $2 \mathcal{A}_1 = \mathcal{A}_2$, $4 \mathcal{A}_3 = 2 \mathcal{A}_{10} = 2 \mathcal{A}_{11} = \mathcal{A}_{14}$, $2 \mathcal{A}_6 = \mathcal{A}_8$, $2 \mathcal{A}_7 = \mathcal{A}_9$, $2 \mathcal{A}_{16} = 4 \mathcal{A}_{17} = \mathcal{A}_{18}$. As a consequence of these relations some terms in (\ref{SET}) become redundant which we take into account by putting $a_1 = a_3 = a_6 = a_7 = a_{10} = a_{11} = a_{16} = a_{17} = 0$. Analogously, we have to consider the derivatives of the gauge condition (\ref{harmonicgauge}) and thus replace the terms of type $h{_{ab}}^{,bc}$ by $\frac{1}{2} h{_{,a}}^{c}$.

Employing the linearized harmonic gauge in the field equations implies a Ricci tensor equal to $R_{ab} = - \frac{1}{2}  h{_{ab,c}}^c $, Ricci scalar $R = - \frac{1}{2} h{_{,c}}^c$, and Einstein tensor $ 2 G_{ab} = -h{_{ab,c}}^c + \frac{1}{2} \eta_{ab} h{_{,c}}^c $. Using \textsc{Cadabra} and some simple rearrangements we arrive at a five-parameter tensor with coefficients given by $\alpha_1 = a_2$, $\alpha_2 = a_4 = - a_{12}$, $\alpha_3 = a_9 = -\frac{1}{2} a_{13} = a_{20}$, $\alpha_4 = a_{14}$, $\alpha_5 = a_{15} = -2 a_{19}$, $a_8 = -\alpha_1 - \frac{1}{2} \alpha_2$, $a_{18} = -\frac{1}{4} \left( \alpha_1 + \alpha_3 + 2 \alpha_4 \right)$. Explicitly,
\begin{align}
  T_{ik} = \,\, &	\alpha_1 \left( h_{ik,a} h^{,a} - h_{ia,k} h^{,a} - \frac{1}{4} \eta_{ik} h_{,b} h^{,b}  \right) + \nonumber \\
  				&	\alpha_2 \left( h_{ia,b} h{_k}^{a,b} - h_{ia,b} h{^{ab}}_{,k} - \frac{1}{2} h_{ia,k} h^{,a} \right) + \nonumber \\
  				&	\alpha_3 \bigg( h_{ka,i} h^{,a} - 2 h_{ka,b} h{^{ab}}_{,i} + \eta_{ik} h_{ab,c} h^{bc,a} \nonumber \\
  				&	\qquad \left. - \, \frac{1}{4} \eta_{ik} h_{,b} h^{,b} \right) + \alpha_4 \left( h_{,i} h_{,k} - \frac{1}{2} \eta_{ik} h_{,b} h^{,b}  \right) + \nonumber \\
  				&	\alpha_5 \left( h_{ab,i} h{^{ab}}_{,k} - \frac{1}{2} \eta_{ik} h_{ab,c} h^{ab,c}   \right).
\end{align}

Therefore, the energy-momentum tensors for the linearized gravity with the harmonic gauge condition chosen \textit{ab initio} form a five-parameter system -- hence, with one additional free parameter as compared with the case not involving any gauge condition. The above expression is in general nonsymmetric. By putting $-\frac{1}{2} \alpha_2 - \alpha_1 = \alpha_3$, $-\alpha_2 = -2\alpha_3$, we arrive at the symmetric expressions which form a three-parameter system. Introducing new constant parameters by $\alpha = \frac{1}{2} \alpha_1 = -\frac{1}{2} \alpha_2 = -\alpha_3$, $\beta = \alpha_4$, $\gamma = \alpha_5$, we get the symmetric tensor in the form
\begin{align}
	T_{ik} =& \, \alpha \left( 2 h_{ik,a} h^{,a} - h_{ia,k} h^{,a} - h_{ka,i} h^{,a} - \frac{1}{4} \eta_{ik} h_{,b} h^{,b}  \right. \nonumber \\
	& \qquad  - 2 h_{ia,b} h{_k}^{a,b} + 2 h_{ia,b} h{^{ab}}_{,k} + 2 h_{ka,b} h{^{ab}}_{,i}  \nonumber \\
	& \qquad \, - \eta_{ik} h_{ab,c} h^{bc,a} \bigg) + \beta \left( h_{,i} h_{,k} - \frac{1}{2} \eta_{ik} h_{,b} h^{,b} \right) + \nonumber \\
	& \gamma \left( h_{ab,i} h{^{ab}}_{,k} - \frac{1}{2} \eta_{ik} h_{ab,c} h^{ab,c} \right).
\end{align}
The tensor suggested in \cite{butcher} follows after choosing $\alpha = 0$, $\beta = -\frac{1}{8}$, $\gamma = \frac{1}{4}$. Hence, our procedure based just on the linear gravity and harmonic gauge shows how the energy-momentum tensor introduced by Butcher \textit{et al}. \cite{butcher}, \cite{butcher2} is contained in a larger (three-parameter) family of conserved symmetric tensors.  Accepting the physical arguments presented in \cite{butcher}, \cite{butcher2}, we arrive at the unique expression.  

  Hence, our procedure shows that the energy-momentum tensor introduced in \cite{butcher} based on the linearized gravity and harmonic gauge is not unique.

It is worth to emphasize that starting from the unique symmetric energy-momentum tensor (\ref{symetrickybicak}) derived \textit{without any gauge condition} we do not arrive at the tensor proposed in \cite{butcher} if we apply the harmonic gauge condition in the expression (\ref{symetrickybicak}) \textit{a posteriori}. In their most recent work, Butcher \textit{et al.} \cite{butcher2} rederive their symmetric expression
\begin{equation}
	8 T_{ik} = - h_{,i} h_{,k} + 2 h_{ab,i} h{^{ab}}_{,k} + \frac{1}{2} \eta_{ik} h_{,b} h^{,b} - \eta_{ik} h_{ab,c} h^{ab,c}  \label{butcher01}
\end{equation}
found in the harmonic gauge from a variational formulation not involving a special gauge condition. They arrive at the result [see (13a) in \cite{butcher2}]
\begin{align}
	4 T_{ik} = \,\,& - 2 h_{ka,i} h{^{ab}}_{,b} + h_{ka,i} h^{,a} + h{_{ka}}^{,a} h_{,i} - h_{,i} h_{,k} \nonumber \\
	& + h_{ab,i} h{^{ab}}_{,k} - \eta_{ik} h_{,b} h{^{bc}}_{,c} + \eta_{ik} h{_{ab}}^{,a} h{^{bc}}_{,c} \nonumber \\
	& + \frac{1}{2} \eta_{ik} h_{,b} h^{,b} - \frac{1}{2} \eta_{ik} h_{ab,c} h^{ab,c}, \label{butcher02}
\end{align}
which under the harmonic gauge condition turns into their original result (\ref{butcher01}). Notice that (\ref{butcher02}) is not symmetric. It is contained in our general form (\ref{gauge1general}): we obtain (\ref{butcher02}) by putting $- 2\alpha_1 = \alpha_2 = 2\alpha_3 = -\frac{1}{2}$ and $\alpha_4 = 0$ in (\ref{gauge1general}).

A general superpotential for linearized gravity in the harmonic gauge reads as follows:
\begin{align}
  U_{ikl} = \,\, & \alpha_1 \left( h_{a[k} h{^{a}}_{l],i} + \frac{1}{2} \eta_{i[k} h_{l]a} h^{,a} - \eta_{i[k} h^{ab} h_{l]a,b} \right) + \nonumber \\
  & 2 \alpha_2 \left( h \, h_{i[k,l]} - \frac{1}{2} \eta_{i[k} h \, h_{,l]} \right) + \nonumber \\
  & 2 \alpha_3 \left( - \frac{1}{2} h_{i[k} h_{,l]} + h_{ia} h{^{a}}_{[k,l]} \right).
\end{align}
Hence, it can be obtained directly from (\ref{superbicak}) by imposing the harmonic gauge condition.

\subsubsection{Generalized gauge condition \label{chgaugeparam}}

The authors of \cite{butcher} consider also the generalized gauge condition of the form $h^{ab}_{,b} = \chi h^{,a}$, where $\chi$ is a constant parameter, which may be called a generalized (or parametrized) harmonic condition. We wish to apply our method also in this more general case. The resulting Ricci and Einstein tensors now read $2 R_{bc} = (2\chi-1) \partial_{bc} h - \partial{_a}^a h_{bc}$ and $2 G_{bc} = (2\chi - 1) \partial_{bc} h - \partial{_a}^a h_{bc} - \eta_{bc} (\chi -1) \partial{_a}^a h$.
We follow the same procedure as in Sec. \ref{chgauge}. Recalling the consequences of the gauge condition applied analogously as before, we find that $a_1$, $a_3$, $a_6$, $a_7$, $a_{10}$, $a_{11}$, $a_{16}$ and $a_{17}$ vanish. Next, we multiply the field equations by Lagrange multipliers, write down the master equation, and employ \textsc{Cadabra}. Observing the results we can easily eliminate a number of Lagrange multipliers except for $\lambda_4$ [cf. (\ref{lambdaprispevek})]. Also, we find very simple relations for the following constants: $a_5 = 0$, $a_4 = -a_{12}$, $a_{13} = -2a_{20}$, $a_{15} = -2a_{19}$. The remaining parameters entering the problem have to satisfy four linear equations:
\begin{align}
	0 = \, & \chi a_2 + \chi a_9 + a_{14} + 2 a_{18} + a_8(2\chi -1), \nonumber \\
	0 = \, & a_2 + a_8 + a_{12}(\chi - 1), \nonumber \\
	0 = \, & a_{14} + (\chi -1)\left[ -a_8 + 2 \lambda_4 + 2a_{19} - \chi a_{12} \right], \nonumber \\
	0 = \, & a_9 - 2\chi a_{20} -2 a_{19}(2\chi -1). \label{lambdaeqn}
\end{align}
Considering first $\chi = 1$, the solution is simple: $a_9 = 2 ( a_{19} + a_{20} )$, $ a_2 = -a_8$, $a_{14} = 0$, $a_{9} = -2a_{18}$. Introducing now four parameters $\alpha_i$ and using the system (\ref{lambdaeqn}), we find $\alpha_1 = a_8 = -a_2$, $\alpha_2 =  a_{12} = -a_4$, $\alpha_3 = a_{19} = -\frac{1}{2} a_{15}$, $\alpha_4 = a_{20} = -\frac{1}{2} a_{13}$, $a_9 = - 2 a_{18} = 2\alpha_3 + 2\alpha_4$. The conserved energy-momentum tensor acquires the following form
\begin{align}
	T_{ik} = \, & \alpha_1 \left(-h_{ik,a} h^{,a} + h_{ia,k} h^{,a} \right) + \alpha_2 \left( - h_{ia,b} h{_k}^{a,b} \right. \nonumber \\
	& \left. + h_{ia,b} h{^{ab}}_{,k} \right) + \alpha_3 \left( 2 h_{ka,i} h^{,a}  - 2 h_{ab,i} h{^{ab}}_{,k}  \right. \nonumber \\
	& \left.  - \, \eta_{ik} h_{,b} h^{,b} + \eta_{ik} h_{ab,c} h^{ab,c} \right) + \alpha_4 \big( 2 h_{ka,i} h^{,a} \nonumber \\ 
	& \left. - \, 2 h_{ka,b} h{^{ab}}_{,i} - \eta_{ik} h_{,b} h^{,b} + \eta_{ik} h_{ab,c} h^{bc,a} \right). \label{sollambda1}
\end{align}
The requirement of symmetry leads to the conditions $\alpha_1 = 2 \alpha_3 + 2 \alpha_4$ and $\alpha_2 = -2\alpha_4$; i.e. it leaves us with a two-parameter system.

For $ \chi \neq 1$, the system of equations (\ref{lambdaeqn}) has the following solution
\begin{align}
	a_9 = \, & 2(2\chi -1) a_{19} + 2\chi a_{20} \label{gaugeL1}, \\
	a_2 = \, & -a_8 + (1-\chi) a_{12} \label{gaugeL2}, \\
	a_{14} = \, & (1-\chi) a_8 + \chi (\chi - 1) a_{12} - 2a_{18} \nonumber \\
	& - 2\chi(2\chi-1) a_{19} -2\chi^2 a_{20} \label{gaugeL3}.
\end{align}
Notice that the third equation in the system (\ref{lambdaeqn}) can just be used to express the multiplier $\lambda_4$ and does not restrict the form of the energy-momentum tensor. Let us now introduce five parameters as follows $\alpha_1 = a_8$, $\alpha_2 = a_{12} = - a_4$, $\alpha_3 = a_{18}$, $\alpha_4 = a_{19} = -\frac{1}{2} a_{15}$, $\alpha_5 = a_{20} = -\frac{1}{2} a_{13}$. Collecting all the previous results for the coefficients $a_i$ [regarding also Eqs. (\ref{gaugeL1}), (\ref{gaugeL2}) and (\ref{gaugeL3})] we find the following expression for the energy-momentum tensor when a generalized harmonic gauge condition is used:
\begin{align}
	T_{ik} = & \, \alpha_1 \left( - h_{ik,a} h^{,a} + h_{ia,k} h^{,a} + (1-\chi) h_{,i} h_{,k} \right) + \nonumber \\
			 & \, \alpha_2 \left( (1-\chi) h_{ik,a} h^{,a} - h_{ia,b} h{_k}^{a,b} + h_{ia,b} h{^{ab}}_{,k}  \right. \nonumber \\
			 & \, \left. \qquad + \chi(\chi - 1) h_{,i} h_{,k}   \right) + \nonumber \\
			 & \, \alpha_3 \left( - 2 h_{,i} h_{,k} + \eta_{ik} h_{,b} h^{,b}  \right) + \nonumber \\
			 & \, \alpha_4 \left( 2(2\chi-1) h_{ka,i} h^{,a} + 2\chi(1-2\chi) h_{,i} h_{,k} \right. \nonumber \\
			 & \, \quad \left. - \, 2 h_{ab,i} h{^{ab}}_{,k} + \eta_{ik} h_{ab,c} h^{ab,c}  \right) + \nonumber \\
			 & \, \alpha_5 \left( 2\chi h_{ka,i} h^{,a} - 2 h_{ka,b} h{^{ab}}_{,i} + \eta_{ik} h_{ab,c} h^{bc,a} \right. \nonumber \\
			 & \, \quad \left.  - \, 2\chi^2 h_{,i} h_{,k}  \right). \label{chitensor}
\end{align}
The requirement of symmetry yields conditions $ \alpha_1 = 2(2\chi - 1) \alpha_4 + 2\chi \alpha_5$ and $\alpha_2 = -2\alpha_5$, so (\ref{chitensor}) becomes a three-parameter system.

The resulting expression (\ref{chitensor}) is meaningful also for $\chi \rightarrow 1$; however, we obtain the solution (\ref{sollambda1}) for $\chi = 1$ after choosing $\alpha_4 = -2\alpha_3$. The three-parameter system of symmetric tensors for $\chi \neq 1$ then goes over to the two-parameter system.

Imagine we demand the independence of the result (\ref{chitensor}) on the parameter $\chi$, i.e., we require the same conserved tensor for any $\chi$. There are three terms that are $\chi$ dependent: $h_{,i} h_{,k}$, $h_{ik,a} h^{,a}$, and $h_{ka,i} h^{,a}$. Writing out explicitly the corresponding part of the energy-momentum tensor we find
\begin{align}
	T_{ik} = \,\, & \left[ (\alpha_1 - 2\alpha_3) + \chi(-\alpha_1-\alpha_2+2\alpha_4) \right. \nonumber \\ 
	& \left. \quad + \chi^2 (\alpha_2-4\alpha_4-2\alpha_5) \right] h_{,i} h_{,k} \, + \nonumber \\
	& \left[ (\alpha_2-\alpha_1) + \chi(-\alpha_2) \right] h_{ik,a} h^{,a} \, + \nonumber \\
	& \left[ (-2\alpha_4) + \chi(4\alpha_4 + 2\alpha_5) \right] h_{ka,i} h^{,a} + \ldots
\end{align}

Therefore, the resulting energy-momentum tensor will be independent of $\chi$ if the coefficients satisfy $\alpha_2 = 0$, $\alpha_1 = 2 \alpha_4$, $\alpha_5 = -2\alpha_4$, forming thus a two-parameter system. This tensor cannot be made symmetric. Finally, adding the condition that the $\chi$-independent tensor is conserved also for $\chi = 1$, i.e. $\alpha_4 = -2 \alpha_3$, we obtain \textit{a unique nonsymmetric} energy-momentum tensor in linearized gravity with parametrized gauge condition $h^{ab}_{,b} = \chi h^{,a}$ which is conserved for arbitrary $\chi$. It reads
\begin{align}
	T_{ik} = \,\, & \alpha \, \Big(-2 h_{ik,a} h^{,a} + 2 h_{ia,k} h^{,a} - 2 h_{ka,i} h^{,a}  \nonumber \\
	 & \quad + 3 h_{,i} h_{,k} - \frac{1}{2} \eta_{ik} h_{,b} h^{,b} - 2 h_{ab,i} h{^{ab}}_{,k}    \nonumber \\
		& \quad + \eta_{ik} h_{ab,c} h^{ab,c} +  4 h_{ka,b} h{^{ab}}_{,i} - 2 \eta_{ik} h_{ab,c} h^{bc,a} \Big).
\end{align}

\subsubsection{High-frequency waves}

In the physically most important case of high-frequency waves propagating in vacuum, the quantities quadratic in $h_{ik,l}$ become gauge invariant after being averaged suitably. This result goes back to the seminal work by Isaacson \cite{isaacson1}, \cite{isaacson2} which entered also classical textbooks; see \cite{MTW}, \cite{landau}, for example. In general, the condition requires the characteristic wavelength to be short compared to the background curvature of spacetime. This is easily satisfied in the linear theory when the background is flat. The ``Brill-Hartle averaging'' is the appropriate technique of constructing the average of an oscillating tensor field in a general background. (In flat backgrounds, one can just average over one period of oscillation in time and one wavelength of distance in spatial directions; see \cite{schutz}, p. 254.) Under the change of gauge, $x \rightarrow x' = x + \xi$, the perturbation
$h \rightarrow h' = h + \partial\xi$, so
\begin{equation} 
(\partial h') (\partial h') \rightarrow (\partial h) (\partial h) + (\partial h)(\partial^2 \xi) + (\partial^2 \xi)(\partial^2 \xi),
\end{equation}
but the last two terms are negligible after averaging. Moreover, since the averaging makes divergences small, we may convert various products of $(\partial h) (\partial h)$ into other terms. For example, 
\begin{equation}
h{_k}^{a,b} h_{bi,a} = - h{_{ka}}^{,ba} h_{bi} + (h{_k}^{a,b} h_{bi})_{,a},
\end{equation}
so after averaging and choosing the gauge with $h{^{ab}}_{,b} = 0$ (see below) this term drops out. In addition, in the curved backgrounds in the high-frequency approximation the covariant derivatives commute (see \cite{isaacson2}, Sec. 4 and the Appendix there for the details).

Regarding these results, it is clear that after averaging, we may omit the divergence of the superpotential in our general energy-momentum tensor (\ref{gauge1generalsuperpot}) in the linear gravity. In addition, since the averaging makes the resulting expressions gauge invariant we may choose a simple gauge. Assuming that we are in a vacuum region we may choose the Lorenz gauge in which $h{^{ab}}_{,b} = 0$ and $h{^a}_a = 0$ so that the harmonic gauge condition (\ref{harmonicgauge}) is automatically satisfied.  Then the terms involving $h$ in (\ref{gauge1generalsuperpot}) drop out, and rewriting the fourth and last two terms in (\ref{gauge1generalsuperpot}) in the way indicated above and using the Lorenz gauge, we arrive at the following simple expression:
\begin{equation}
\langle T_{ik} \rangle =  \mbox{const} \cdot \langle h_{ab,i} h{^{ab}}_{,k} \rangle, \label{averagedTik}                                                     
\end{equation}
where the brackets $\langle \, \rangle$ denote the averaging; the same expression follows from the symmetric tensor (\ref{gauge1generalsuperpotsym}). And it is easy to see that the averaged energy-momentum tensor introduced by Butcher \textit{et al} \cite{butcher}, \cite{butcher2} leads to exactly the same result. In fact, even in the case of a curved vacuum spacetime the averaging of the ``metric energy-momentum tensor'' (\ref{subsubvaritensor}) in the the generalized Lorenz gauge $h{^{ab}}_{;b} = 0$, $h{^a}_a = 0$ implies (\ref{averagedTik}) with partial derivatives replaced by covariant ones.

\section{Massive gravity \label{massivegravitychapter}}

Finally, we turn to the case of the massive gravity in a vacuum. We start from the Fierz-Pauli action for the massive gravity (a massive spin-2 particle -- see, for example, \cite{bichler}) described by symmetric tensor $h_{ab}$: 
\begin{align}
  S_{FP} &= \int \Big[ -\frac{1}{2} h_{ab,c} h^{ab,c} + h_{ab,c} h^{bc,a} - h{^{ab}}_{,a} h_{,b} \nonumber \\
  & \qquad + \frac{1}{2} h_{,a} h^{,a} - \frac{1}{2} m^2 \left( h_{ab} h^{ab} - h^2 \right) \Big] d^4x. \label{FPaction}
\end{align}
The equations of motion following from this action have the form
\begin{align}
  \frac{\delta S}{\delta h^{ab}} = \,\,& h{_{ab,c}}^c - h{_{ac,b}}^c - h{_{bc,a}}^c + \eta_{ab} h{_{cd}}^{,cd} + h_{,ab} \nonumber \\
  & - \eta_{ab} h{_{,c}}^{c} - m^2 \left( h_{ab} - \eta_{ab} h \right) = 0. \label{FPeqn}
\end{align}
The divergence of the last equation with respect to a free index implies, for $m \neq 0$, $h{_{ab}}^{,b} - h_{,a} = 0$. Substituting back into (\ref{FPeqn}) and making contraction in free indices we find that the trace $h$ has to vanish, $h = 0$. Equations (\ref{FPeqn}) are thus equivalent to the following set of equations:
\begin{equation}
   h{_{ab,c}}^c - m^2 h_{ab} = 0, \qquad h{_{ab}}^{,b} = 0, \qquad h = 0. \label{FPeqns}
\end{equation}

\subsection{Klein-Gordon equation \label{subKGeqn}}

 Starting first just with the Klein-Gordon equation,
\begin{equation}
	h{_{ab,c}}^c - m^2 h_{ab} = 0,
\end{equation}
we obtain the following five-parameter result for conserved tensors: $a_1 = a_3 = a_5 = a_9 = a_{11} = a_{16} = 0$, $\alpha_1 = a_7 = -a_{13} = -2a_{17} = 2a_{20}$, $\alpha_2 = \frac{1}{m^2} c_1 = a_2 = - a_8 = - a_{10}$, $\alpha_3 = \frac{1}{m^2} c_2 = a_4 = - a_6 = - a_{12}$, $\alpha_4 = \frac{1}{m^2} c_3 = - \frac{1}{2} a_{14} = a_{18}$, $\alpha_5 = \frac{1}{m^2} c_4 = - \frac{1}{2} a_{15} = a_{19}$, where the meaning of the constants $c_i$ is explained in (\ref{additionalSET}) and (\ref{generalizedSET}). The explicit expression for the energy-momentum tensor looks as follows:
\begin{align}
  T_{ik} = \,\,	& \alpha_1 \, T_{ik}^{(strong)} + \nonumber \\
				& \alpha_2 \left( m^2 h_{ik} h + h_{ik,a} h^{,a} - h_{ia,k} h^{,a} - h{_{ia}}^{,a} h_{,k} \right) + \nonumber \\
				& \alpha_3 \left( m^2 h_{ia} h{_k}^a + h_{ia,b} h{_k}^{a,b} - h_{ia,k} h{^{ab}}_{,b} \right. \nonumber \\
				& \qquad \left. - \, h_{ia,b} h{^{ab}}_{,k} \right) + \nonumber \\
				& \alpha_4 \left( m^2 \eta_{ik} h^2 - 2 h_{,i} h_{,k} + \eta_{ik} h_{,b} h^{,b} \right) + \nonumber \\
				& \alpha_5 \left( m^2 \eta_{ik} h_{ab} h^{ab} - 2 h_{ab,i} h{^{ab}}_{,k} + \eta_{ik} h_{ab,c} h^{ab,c} \right). \label{KGvysledek}
\end{align}
The five-parameter system (\ref{KGvysledek}) reduces just to a two-parametric one with $\alpha_1 = \alpha_2 = \alpha_3 = 0$ if we require the energy-momentum tensor to be symmetric.

Applying the additional conditions $h=0$ and $h{_{ab}}^{,a}=0$ on the resulting expression (\ref{KGvysledek}) we arrive at
\begin{align}
	T_{ik} = \,\,& \alpha_1 \left( - h_{ka,b} h{^{ab}}_{,i} + \frac{1}{2} \eta_{ik} h_{ab,c} h^{bc,a} \right) + \nonumber \\
			& \alpha_3 \left( m^2 h_{ia} h{_k}^a + h_{ia,b} h{_k}^{a,b} - h_{ia,b} h{^{ab}}_{,k} \right) + \nonumber \\
			& \alpha_5 \left( m^2 \eta_{ik} h_{ab} h^{ab} - 2 h_{ab,i} h{^{ab}}_{,k} + \eta_{ik} h_{ab,c} h^{ab,c} \right). \label{massiveKGuh}
\end{align}
The requirement of symmetry implies $\alpha_3 = \alpha_1$, which leads to the following expression
\begin{align}
	T_{ik} =\,\,& \alpha_1 \bigg( m^2 h_{ia} h{_k}^a + h_{ia,b} h{_k}^{a,b} - h_{ia,b} h{^{ab}}_{,k} \nonumber \\
	& \left. \qquad - \, h_{ka,b} h{^{ab}}_{,i} + \frac{1}{2} \eta_{ik} h_{ab,c} h^{bc,a}  \right) + \nonumber \\
			& \alpha_5 \left( m^2 \eta_{ik} h_{ab} h^{ab} - 2 h_{ab,i} h{^{ab}}_{,k} + \eta_{ik} h_{ab,c} h^{ab,c} \right).
\end{align}
By choosing $\alpha_1 = 1$, $\alpha_5 = -\frac{1}{4}$, we obtain the ``generalized'' linearized Landau-Lifshitz pseudotensor:
\begin{align}
	T_{ik}^{(LL)} = \,\,& \frac{1}{2} h_{ab,i} h{^{ab}}_{,k} - \frac{1}{4} \eta_{ik} h_{ab,c} h^{ab,c} + \frac{1}{2} \eta_{ik} h_{ab,c} h^{bc,a} \nonumber \\
	& + h_{ia,b} h{_k}^{a,b} - h_{ia,b} h{^{ab}}_{,k} - h_{ka,b} h{^{ab}}_{,i} \nonumber \\
	& + m^2 \left( h_{ia} h{_k}^a - \frac{1}{4} \eta_{ik} h_{ab} h^{ab} \right). \label{LLmassive}
\end{align}
Putting $m=0$ we recover the symmetric energy-momentum tensor of the Einstein linearized theory (\ref{symetrickybicak}) after we substitute therein the second and the third condition in Eq. (\ref{FPeqns}); i.e. we obtain the standard Landau-Lifshitz pseudotensor (\ref{LLtensor}) linearized and with these two conditions taken into account.

If we use the same procedure as in Secs. \ref{chgauge} and \ref{chgaugeparam}, i.e., we first apply the equations $h{^{ab}}_{,b} = 0$ and $h=0$ in the general form of energy-momentum tensor (\ref{additionalSET}) and (\ref{generalizedSET}), only nonvanishing terms are then $\mathcal{A}_{4}$, $\mathcal{A}_{5}$, $\mathcal{A}_{12}$, $\mathcal{A}_{13}$, $\mathcal{A}_{15}$, $\mathcal{A}_{19}$, $\mathcal{A}_{20}$, $\mathcal{C}_{2}$ and $\mathcal{C}_{4}$\footnote{The terms vanishing due to the equations $h{^{ab}}_{,b} = h = 0$ can be added with any coefficient to the resulting tensor, but if the above field equations are satisfied the energy-momentum tensor does not, of course, change.}. The resulting three-parameter energy-momentum tensor is again (\ref{massiveKGuh}).

\subsection{A unique symmetric energy-momentum tensor from the Fierz-Pauli equation \label{subFPunique}}

Finally, starting from the field equation (\ref{FPeqn}) and general form of energy-momentum tensor (\ref{additionalSET}) and (\ref{generalizedSET}), we find that the tensor is conserved modulo the Fierz-Pauli equation (\ref{FPeqn}) provided that the following relations between the corresponding nonvanishing coefficients are satisfied: $\alpha_1 = a_7 = -2a_{17}$, $\alpha_2 = a_9 = a_{11} = -a_{14} = a_{15} = -a_{16} = 2a_{18} = -2a_{19} = \frac{2}{m^2} c_{3} = -\frac{2}{m^2} c_{4}$, $a_{13} = -\alpha_1 - 2\alpha_2$, $a_{20} = \frac{1}{2} \alpha_1 + \alpha_2$. 
These relations lead to the following explicit form of the energy-momentum tensor:
\begin{align}
	T_{ik} =\,\,& \alpha_1 T_{ik}^{(strong)} + \alpha_2 \bigg( h_{ka,i} h^{,a} + h{_{ka,}}^a h_{,i}  \nonumber \\
	& - 2 h_{ka,b} h{^{ab}}_{,i} - h_{,i} h_{,k} + h_{ab,i} h{^{ab}}_{,k} - \eta_{ik} h_{,b} h{^{bc}}_{,c} \nonumber \\
	& + \frac{1}{2} \eta_{ik} h_{,b} h^{,b} - \frac{1}{2} \eta_{ik} h_{ab,c} h^{ab,c} + \eta_{ik} h_{ab,c} h^{bc,a} \nonumber \\
	& + \frac{1}{2} m^2 \eta_{ik} h^2 - \frac{1}{2} m^2 \eta_{ik} h_{ab} h^{ab}  \bigg).  \label{tenzorFP} 
\end{align}
Notice that this result, after putting $m=0$, coincides with the part of the energy-momentum tensor for the linearized gravity (\ref{gauge1general}). However, to see it, we must, because of a different parametrization, make the change $\alpha_1 \rightarrow \alpha_2$, $\alpha_2 \rightarrow \alpha_1$, $\alpha_3 \rightarrow -\alpha_2$, and $\alpha_4 \rightarrow 0$ in (\ref{gauge1general}); then (\ref{tenzorFP}) follows. It is noteworthy to observe that the inclusion of massive terms \textit{reduces} the nonuniqueness of resulting conserved tensors.

Curiously enough, the energy-momentum tensor conserved as a consequence of the Fierz-Pauli equation in its original form (\ref{FPeqn}) cannot be made symmetric for any choice of parameters $\alpha_1$, $\alpha_2$. However, applying differential operations on the original Fierz-Pauli equation (which give rise to the appearance of the third derivatives) we know that Eqs.(\ref{FPeqns}) are implied. Using the second and the third equation of (\ref{FPeqns}) the tensor (\ref{tenzorFP}) then turns into the following expression
\begin{align}
	\tilde{T}_{ik} = \,\, & \alpha_1 \left( - h_{ka,b} h{^{ab}}_{,i} + \frac{1}{2} \eta_{ik} h_{ab,c} h^{bc,a} \right) + \nonumber \\
					    & \alpha_2 \left( - 2 h_{ka,b} h{^{ab}}_{,i} + h_{ab,i} h{^{ab}}_{,k} - \frac{1}{2} \eta_{ik} h_{ab,c} h^{ab,c} \right. \nonumber \\
					   & \left. \qquad + \, \eta_{ik} h_{ab,c} h^{bc,a} - \frac{1}{2} m^2 \eta_{ik} h_{ab} h^{ab}    \right). \label{massivefinal2}
\end{align}
This tensor can be made symmetric by the choice $\alpha = \alpha_2 = -\frac{1}{2} \alpha_1 $ obtaining thus \textit{a unique} symmetric tensor for linear massive gravity in the form
\begin{equation}
	\bar{T}_{ik} = \alpha \left( h_{ab,i} h{^{ab}}_{,k} - \frac{1}{2} \eta_{ik} h_{ab,c} h^{ab,c} - \frac{1}{2} m^2 \eta_{ik} h_{ab} h^{ab}  \right). \label{massivefinal3}
\end{equation}
Observe that the resulting unique symmetric tensor does not coincide with the linearized Landau-Lifshitz pseudotensor generalized to massive gravity. It is simpler.

It is interesting to compare the expressions (\ref{tenzorFP})--(\ref{massivefinal3}) with the standard results following from the variational principle and Noether's theorem. With the Lagrangian density $L$ determined by the Fierz-Pauli action (\ref{FPaction}) (with a multiplicative constant omitted), the \textit{canonical} energy-momentum tensor
\begin{equation}
	T{_i}^{k\,(can)} = L \, \delta{_i}^k - \frac{\partial L}{\partial h_{ab,k}} h_{ab,i},
\end{equation}
turns out to be exactly the expression multiplied by $\alpha_2$ in (\ref{tenzorFP}). Substituting then from the second and the third equation of (\ref{FPeqns}) as before, we get
\begin{align}
	T_{ik}^{(can)} = \, & h_{ab,i} h{^{ab}}_{,k} - 2 h_{ab,i} h{_k}^{a,b} + \eta_{ik} \left( -\frac{1}{2} h_{ab,c} h^{ab,c} \right. \nonumber \\
		& \left. + \, h_{ab,c} h^{bc,a} - \frac{1}{2} m^2 h_{ab} h^{ab} \right).
\end{align}
Therefore, Eq. (\ref{massivefinal2}) can be written in the form
\begin{equation}
	\tilde{T}_{ik} = \alpha_1 T_{ik}^{(strong)} + \alpha_2 T_{ik}^{(can)}.
\end{equation}
Putting then $\alpha_2 = -\frac{1}{2} \alpha_1$ we arrive at (\ref{massivefinal3}). Since the first, strongly conserved part can be derived from the superpotential (\ref{strongsuperpotential}), the total quantities can be evaluated by using just $T_{ik}^{(can)}$. The same total quantities will, of course, result also from the uniquely given symmetric tensor (\ref{massivefinal3}). The ``metric energy-momentum tensor'' following from the variational principle by the same procedure as the expression (\ref{subsubvaritensor}) was obtained, contains the second derivatives $\partial^2 h$. The Belinfante procedure of the symmetrization (i.e. the metric energy-momentum tensor) in the case of higher spin fields gives rise to new types of contributions to energy-momentum tensors, in our case $\propto h \, \partial^2 h$, absent in the lower spins. Our method of a systematic construction of superpotentials enabled us to find such an expression which makes the canonical tensor symmetric and the tensor involves fields and their first derivatives only. The unique expression (\ref{massivefinal3}) following from the Fierz-Pauli equation (action) is thus to be preferred. Putting $m = 0$ and $\alpha = \frac{1}{4}$ in (\ref{massivefinal3}), we arrive at the tensor (\ref{butcher01}) advocated in \cite{butcher}, with $h = 0$.

Finally, let us note that our simple symmetric tensor (\ref{massivefinal3}) differs from the Landau-Lifshitz tensor (\ref{LLmassive}) by the divergence of a superpotential; hence, both expressions lead to the same total (integrated) quantities provided that the field falls off appropriately at infinity. Regarding the superpotential (\ref{strongsuperpotential}) -- which leads to the strongly conserved tensor -- and puting $h{^{ab}}_{,b} = 0$ and $\alpha = -1$, it reads
\begin{equation}
	U_{ikl} = \eta_{i[k} h^{ab} h_{l]a,b} - h_{a[k} h{^a}_{l],i}.
\end{equation}    
Introduce then another superpotential
\begin{equation}
	\bar{U}_{ikl} = 2 h_{ia} h{^a}_{[k,l]},
\end{equation}
and use the first two field equations in (\ref{FPeqns}) when evaluating its divergence. As a result we find that
\begin{equation}
	T_{ik}^{(LL)} = \bar{T}_{ik} + \left( U{_{ik}}^l + \bar{U}{_{ik}}^l \right)_{,l},
\end{equation}
where $T_{ik}^{(LL)}$ is given by (\ref{LLmassive}) and $\bar{T}_{ik}$ by (\ref{massivefinal3}) with $\alpha = \frac{1}{2}$.

\section{Acknowledgements}

J.B. acknowledges the support from the Czech Science Foundation, GA\v{C}R Grant No. 14-37086G (Albert Einstein Centre); J.S. was supported by the Grant Agency of the Czech Technical University in Prague, Grant No. SGS13/217/OHK4/3T/14. We also thank for the hospitality of Albert Einstein Institute in Golm where we enjoyed brief but useful collaboration.

\appendix

\section{Cadabra \label{cadabra}}

\textsc{Cadabra} is a computer algebra system designed for solving the problems in field theory (see \cite{cadabra1}, \cite{cadabra2}). Here we used its effectiveness in manipulating complicated tensor expressions. In particular with \textsc{Cadabra} software it is easy to obtain equations for multiplicative coefficients $a_i$ (and $c_i$, $\lambda_i$,...) at specific covariant terms. In our case this would be a very tedious task because of the overwhelming number of terms. In \textsc{Cadabra} each term has to be converted into its ``canonical'' form\footnote{The concrete appearance of every term depends on internal working of \textsc{Cadabra} algorithms and the way of storing tensorial structures.}. Grouping the terms and collecting their coefficients generates a set of linear equations as coefficients at each term have to vanish in order to satisfy the master equation (\ref{covariantmaster}).

To illustrate our use of \textsc{Cadabra} we shall briefly describe the code which leads to the resulting energy-momentum tensor (\ref{gauge1general}) of the linearized gravity. We first define tensor indices, metric tensor $g_{ab} = \eta_{ab}$, field variables $h_{ab}$, and its dependence on the partial derivative:
\begin{verbatim}
{a,b,c,d,e,f,i,k,l#}::Indices.
{a,b,c,d,e,f,i,k,l#}::Integer(1..N).

g_{a b}::Metric. g^{a b}::InverseMetric.
g^{a}_{b}::KroneckerDelta.
g_{a}^{b}::KroneckerDelta.

h_{a b}::Symmetric.
\partial_{#}::PartialDerivative.
h_{a b}::Depends(\partial).
\end{verbatim}

The next step is to insert the equation of motion $R_{bc} = 0$ and corresponding Lagrange multipliers forming the right-hand side of the master equation (\ref{covariantmaster}):
\begin{verbatim}
EQM :=   \partial_{b a}{ h^{a}_{c} }
       + \partial_{c a}{ h^{a}_{b} }
       - \partial_{b c}{ h_{a}^{a} }
       - g^{a d} \partial_{a d}{ h_{b c} };

L:=(\lambda_1 g_{i}^{b}
      \partial^{c}{h_{a}^{a}} + ... +
    \lambda_6 \partial_{i}{h^{b c}}) @(EQM);
\end{verbatim}

The following set of \textsc{Cadabra} commands converts all terms into the canonical form:
\begin{verbatim}
@distribute!(%): @eliminate_metric!(%):
@eliminate_kr!(%): @prodsort!(%):
@canonicalise!(%): @rename_dummies!(%);
\end{verbatim}

The last input is the general form of the energy-momentum tensor:
\begin{verbatim}
EMT := A_{1} \partial_{a}{ h_{i k} }
             \partial_{b}{ h^{a b} } + ... +
     + A_{20} g_{i k} \partial_{c}{ h_{a b} }
              \partial^{a}{ h^{b c} };
\end{verbatim}

Now we need to calculate its divergence and convert it to its canonical form to obtain the left-hand side of the master equation:
\begin{verbatim}
divEMT := \partial^{k}{ @(EMT) }:

@distribute!!(%): @prodrule!(%):
@unwrap!(%): @sumflatten!(%): 
@eliminate_metric!(%): @eliminate_kr!(%):
@prodsort!(%): @canonicalise!(%): 
@rename_dummies!(%);
\end{verbatim}

Subtracting the computed terms and collecting the coefficients in front of canonicalized terms leads to the desired linear equations determining the coefficients and thus the conserved tensor:

\begin{verbatim}
@(divEMT) - @(L):

@distribute!(%):
@factor_in!(%){A_{1},...,A_{20},
   \lambda_1,...,\lambda_6};
\end{verbatim}

Finally, the \textsc{Cadabra} output looks explicitly as follows:
\begin{align}
1 := \,\, & ({A}_{1} + {A}_{3} - {\lambda}_{2}) {\partial}^{a}{{h}_{a}\,^{b}}\,  {\partial}_{b}\,^{c}{{h}_{i c}}\,  + \ldots + \nonumber \\
	& ({A}_{15} + {\lambda}_{6}) {\partial}_{i}{{h}^{a b}}\,  {\partial}^{c}\,_{c}{{h}_{a b}}\, ;
\end{align}

\end{document}